\documentclass[apj]{emulateapj}

\newcommand{\myemail}{schenker@astro.caltech.edu}
\newcommand{\lt}{\ifmmode\,<\,\else \,$<$\,\fi}
\newcommand{\kms}{\ifmmode\,{\rm km}\,{\rm s}^{-1}\else km$\,$s$^{-1}$\fi}

\newcommand{\magarc}{\ifmmode {{{{\rm mag}~{\rm arcsec}}^{-2}}}
             \else {{{mag}$~${arcsec}$^{-2}$}}
             \fi}

\newcommand{\bhst}{$B_{435}$}
\newcommand{\vhst}{$V_{606}$}
\newcommand{\ihst}{$i_{775}$}
\newcommand{\zhst}{$z_{850}$}
\newcommand{\yhst}{$Y_{105}$}
\newcommand{\jhst}{$J_{125}$}
\newcommand{\hhst}{$H_{160}$}
\newcommand{\fesc}{$f_{esc}$}

\shorttitle{Nebular Emission and High Redshift Galaxies}

\begin{document}

\title{Contamination of Broad-Band Photometry by Nebular Emission in High Redshift Galaxies: Investigations with
Keck's MOSFIRE Near-Infrared Spectrograph}


\author {Matthew A Schenker\altaffilmark{1},
Richard S Ellis\altaffilmark{1}
Nick P Konidaris\altaffilmark{1},
Daniel P Stark\altaffilmark{2,3},
}

\altaffiltext{1}{Department of Astrophysics, California Institute of 
                Technology, MC 249-17, Pasadena, CA 91125; 
                \myemail}
\altaffiltext{2}{Department of Astronomy and Steward Observatory, University of Arizona, Tucson AZ 85721}
\altaffiltext{3}{Hubble Fellow}


\begin{abstract}

Earlier work has raised the potential importance of nebular emission in the derivation of
the physical characteristics of high redshift Lyman break galaxies. Within certain redshift ranges,
and especially at $z\simeq 6-7$, such lines may be strong enough to reduce estimates of the
stellar masses and ages of galaxies  compared those derived assuming broad-band photometry 
represents stellar light alone. To test this hypothesis at the highest redshifts where such lines 
can be probed with ground-based facilities, we examine the near-infrared spectra of a representative 
sample of 20 $3.0<z<3.8$ Lyman break galaxies using the newly-commissioned MOSFIRE near-infrared 
spectrograph at the Keck I telescope. We use this data to derive the rest-frame equivalent widths (EW) of [O III] 
emission and show that these are comparable to estimates derived using the SED fitting technique
introduced for sources of known redshift by Stark et al (2013). Although our current sample is
modest, its [O III] EW distribution is consistent with that inferred for H$\alpha$ based on SED fitting
of Stark et al's larger sample of $3.8 < z < 5$ galaxies. For a subset of survey galaxies,  we use the combination of
optical and near-infrared spectroscopy to quantify kinematics of outflows in $z\simeq$ 3.5 star-forming galaxies,
and discuss the implications for reionization measurements. The trends we uncover underline the dangers of 
relying purely on broad-band photometry to estimate
the physical properties of high redshift galaxies and emphasize the important role of diagnostic spectroscopy.

\end{abstract}


\section{Introduction}\label{sec:intro}

Detailed photometry of Lyman Break galaxies undertaken with the {\it Hubble Space Telescope (HST)}
and the {\it Spitzer Space Telescope} has provided spectral energy distributions (SEDs) for large
samples of Lyman break galaxies (LBGs) in the important redshift range $3<z<$7. These data have 
been used to derive valuable estimates of the star formation rates (SFRs), stellar masses and ages 
\citep[e.g.,][]{Stark2009a,Labbe2010a,Gonzalez2011a}. The stellar
mass density is a particularly important measure for the epoch just after reionization ended, as it
provides a key indication of whether yet earlier star formation at $z>6$ is capable of maintaining cosmic
reionization \citep{Robertson2013a}.

In an earlier paper \citep{Stark2013a}, we used our large Keck spectroscopic redshift
survey of $3<z<7$ LBGs \citep{Stark2010a,Stark2011a} to investigate earlier claims 
\citep{Schaerer2009a,Schaerer2010a,Ono2010a,deBarros2012a} that emission from rest-frame
optical nebular emission lines (e.g. [O II], [O III], H$\alpha$) may contaminate the broad-band
fluxes uses to derive the above physical quantities. Examining the SEDs of galaxies
of known spectroscopic redshift ( provides a sounder basis for such an investigation compared to  
`forward modeling'  techniques, based on adding nebular emission to stellar population 
synthesis models, which cannot provide an unambiguous result using broad-band data alone.  We used
this technique, pioneered in \cite{Shim2011a} to compare 
SEDs in a redshift range $3.8<z<5.0$ where line contamination is likely, with those in an uncontaminated
range $3.1<z<3.6$, and show that H$\alpha$ emission typically contributes 
30\% of the flux at 3.6$\mu$m. From this comparison we derived a rest-frame equivalent
width distribution for H$\alpha$ which was used to show that nebular emission is likely to have
a significant impact for galaxies at $z\simeq6-7$ where both warm IRAC filters are
contaminated by [O III] and H$\alpha$. 

An additional motivation for the \cite{Stark2013a} study was our interest in addressing
a puzzle which arose from early measures of the specific star formation rate (sSFR) for
$z>2$. It had been claimed that this quantity does not evolve strongly between
$z\simeq2$ and 7 \citep[e.g.,][]{Stark2009a,Gonzalez2010a} in contrast
to numerical simulations \citep[e.g.,][]{Dave2012a} which predict that the sSFR should
closely match the inflow rate of baryonic gas. In \cite{Stark2013a}, we used the equivalent width distribution
of H$\alpha$ derived from the Keck sample to show that the sSFR is likely to evolve 
more rapidly at $z>4$ than previously thought, supporting up to a five-fold increase
between $z\simeq 2$ and 7 (see also \citealt{deBarros2012a}; c.f. \citealt{Gonzalez2012a}).

In this paper, we take the subject to the next logical stage by verifying {\it directly
with near-infrared spectroscopy}, the impact of nebular emission in the analysis
of SEDs for high redshift LBGs. Prior to the launch of the {\it James
Webb Space Telescope (JWST)} it is not possible to directly examine contamination by $H\alpha$
emission within the IRAC warm filters. However, in a manner similar to that employed by
\cite{Stark2013a}, we can investigate contamination by [O III] 5007 \AA\  in the photometric
$K_S$ band at 2.2$\mu$m by studying a representative sample of $3.0<z<3.8$ spectroscopically-confirmed
LBGs. Our goal is to determine the rest-frame equivalent width distribution of [O III]
directly, and to compare this to the extent possible with that inferred for H$\alpha$ from the 
SED-based study of \cite{Stark2013a}. Such a spectroscopic program is made 
possible by the arrival at the Keck 1 telescope of the multi-slit near-infrared spectrograph 
MOSFIRE (McLean et al 2012) which offers the advantage of a significant multiplex gain.
Using this new instrument, we target a representative sample of LBGs selected
to lie within the $3.0<z<3.8$ redshift range in the GOODS North field. We can thus take advantage of 
ACS photometry from the GOODS survey and improved near-infrared photometry from the CANDELS survey.
This extensive photometry further means we can directly compare measured [O III] fluxes
with those inferred using the SED-based approach.

Throughout this paper, we adopt a $\Lambda-$dominated, flat universe with $\Omega_{\Lambda}=0.7$, 
$\Omega_{M}=0.3$ and $\rm{H_{0}}=70\,\rm{h_{70}}~{\rm km\,s}^{-1}\,{\rm Mpc}^{-1}$. 
All magnitudes in this paper are quoted in the AB system \citep{Oke1983a}.  We will refer
to the HST ACS and WFC3/IR filters F435W, F606W, F775W, F850LP, F105W, F125W, and F160W 
as \bhst, \vhst, \ihst, \zhst, \yhst, \jhst, \hhst, respectively.

\section{Target selection}

\subsection{Photometry}

For our target selection and SED fitting, it was necessary to assemble a full
multi-wavelength catalog across the GOODS-N field. For the HST ACS data, we use the publicly available v2.0 GOODS-N
mosaics \citep{Giavalisco2004a}.  For the newly-obtained CANDELS WFC3/IR data
\citep{Grogin2011a, Koekemoer2011a}, we combine the single epoch mosaics publicly available 
as of March 2013 weighting by exposure time, using the image combination routine 
SWARP \citep{Bertin2002a}.  Our CANDELS reductions comprise the first 7 epochs taken in GOODS-N, and
so a number of objects do not yet have coverage in the \yhst\ filter. However,  as we have \jhst\ and \hhst ,
imaging for all targets, this does not constitute a significant weakness for our SED analyses.

To compute accurate colors for each object, we measured the flux of each object using isophotal
apertures determined by SExtractor in PSF-matched images.  Colors were corrected
to a total magnitudes by computing the offset from MAG AUTO for the \ihst\ band in the 
GOODS v2.0 catalog.  In cases where MAG AUTO was deemed to be unreliable by visual inspection 
of the images, we instead used the offset to a flux measured in a 1.0'' diameter aperture to 
derive a total magnitude. 

For the key photometry in the spectral region of interest, we use the $K_S$-band image derived from ultra-deep
Canada France Hawaii Telescope imaging published by \cite{Wang2010a}.  This image has total
exposure time $t = 49.4$ hr and a 5$\sigma$ limiting magnitude of 24.5 in the GOODS-N field.
As the FWHM of the image is 0.7-0.8'', we do not attempt to PSF-match the other images to this coarser resolution.
Rather, as our objects are largely compact, we perform photometry in 1.0'' diameter apertures,
and apply a correction of 0.7 mag for flux falling outside the aperture, determined from analyses of
isolated, unsaturated stellar sources in the $K_S$ image.  Finally, where available, we use deconfused
Spitzer IRAC 3.6 and 4.5 $\mu$m photometry. 

\subsection{Spectroscopic Sample}

Our primary source of targets consisted of spectroscopically-confirmed galaxies from the Keck 
survey of $3<z<6$ LBGs \citep{Stark2010a,Stark2011a} in the GOODS-N field. 
Briefly, this sample was compiled via optical follow-up of color-selected Lyman break galaxies ($B$, $v$, and $i$-drops) 
with the DEIMOS spectrograph on Keck II.  The relevant observations took place between 2008 and 2010, and integration 
times for the sample considered here ranged from 5.0 to 7.0 hours.  The interested reader can find the 
full details of the sample in \cite{Stark2010a}\footnote{A final catalog of
this extensive spectroscopic survey is now being prepared for publication (Stark et al, in prep)}.

\begin{figure*}
\begin{center}

\includegraphics[width=0.6\textwidth]{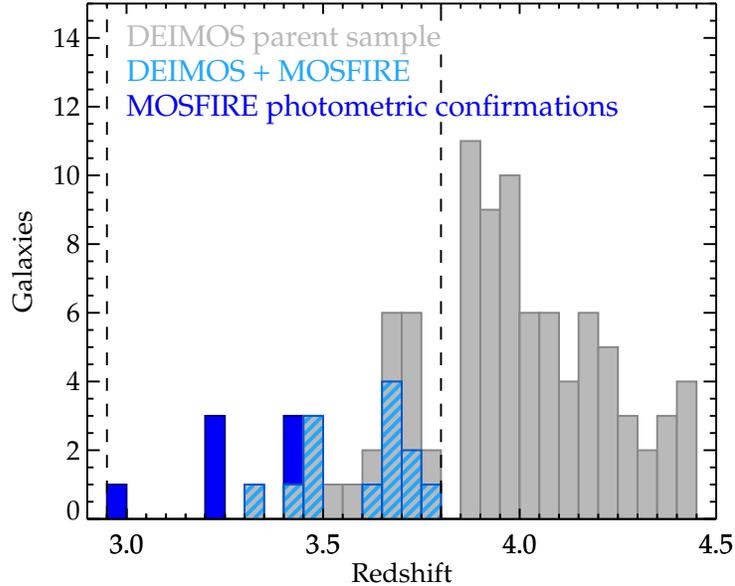}

\caption{\label{fig:zspec_hist} Redshift distribution of Lyman Break Galaxies targeted with MOSFIRE.  The grey histogram indicates
the parent sample within GOODS-N from our prior DEIMOS campaign. The light blue cross-hatched histogram 
denotes the subset of the DEIMOS
spectroscopic sample studied with MOSFIRE and the dark blue histogram
that drawn from a photometric selection (see Section 2). Dashed lines show the boundaries within which [O III] is expected to
contaminate the $K_S$ photometry.}
\end{center}
\end{figure*}

From this compilation, we chose to undertake near-infrared spectroscopy of LBGs, primarily $B$-drops, with 
confirmed redshifts $3.0< z < 3.8$ since both H$\beta$ and [OIII] 4959 + 5007 lie within the 
MOSFIRE K-band transmission window.  As the original $B$-drop sample only sparsely populates the above
redshift range, in anticipation of the present needs, we increased the 
available sample via further DEIMOS observations in June 2012 using photometric redshifts to
improve the redshift coverage. To achieve this, we first created a catalog of \bhst \,dropout galaxies using the 
GOODS v2.0 catalogs and the selection criteria outlined in \cite{Stark2009a}. These SEDs
were then evaluated with a photometric redshift code to assess their chances of lying at $z \leq 3.8$.  
This is a key step, as the $B$-drop sample possesses a mean redshift of $z \sim 4.0$, and only
$\simeq 25 \%$ of $B$-drops lie below $z = 3.8$ (see Fig. \ref{fig:zspec_hist}).  Our priorities for 
inclusion of these targets on the mask were the probability of lying at $z < 3.8$ and 
the \zhst \,magnitude for each target, with brighter targets favored.

This investment of spectroscopic observing time enables us to select targets
that are known {\it a priori} to lie in the accessible redshift range,
and maximizes our efficiency.  Additionally, in the event of non-detections of nebular emission
lines, prior knowledge of the redshift affords robust upper limits on the fluxes.  

Since we are interested in the nebular emission line properties of this sample, it is important to note
that all the targets for the various DEIMOS campaigns were selected only using ACS photometry, 
and thus should not be significantly biased towards objects with strong nebular emission.  

\subsection{Photometric Sample}

As MOSFIRE can accommodate as many as 46 slits on a single mask, we sought to augment our 
spectroscopic sample above with further photometrically-selected LBGs.  Our procedure for
adding new targets was largely as described above, except the SEDs now incorporated
CANDELS WFC3 data where appropriate, and deconfused Spitzer IRAC photometry.  

Although one of the goals of this campaign was to verify the technique pioneered in \cite{Shim2011a} and \cite{Stark2013a}
of using photometric excesses to determine line strengths, we specifically avoided prioritizing targets
by their K-band magnitude.  This allows us to construct an unbiased sample, and thus a better estimate
of the true distribution of nebular line equivalent widths in the following analysis.

\section{Observations}

The targets defined above were observed using MOSFIRE \citep{McLean2012a} on the Keck II telescope.  We observed
two masks on separate observing runs.  The first mask was observed on the night of March 20-21, 2013 
for a total of 4.25 hours integration time, of which $\simeq 1$ hour was affected by thin cloud.  The average seeing for
this mask was 0.75 arcseconds full-width at half maximum (FWHM).  Our second mask was observed on the night of 
Apr 16-17, 2013.  We obtained a total of 2.50 hours of useful integration time, with approximately 40 minutes
of thin clouds, and an average seeing of 0.70 arcseconds FWHM.

We used an ABAB dither pattern with individual exposures of 180 seconds and a slit width of 0.7 arcseconds 
for all targets.  On each mask, in addition to our science targets, we also included a star with 
$K_{AB} < 19.0$ to use for flux calibration.  This ensures accurate accounting for slit loss due to seeing and 
pointing errors, as the star is observed in exactly the same manner as all targets.  

The data was reduced using the publicly available MOSFIRE data reduction pipeline\footnote{\rm https://code.google.com/p/mosfire/}.  
Briefly, the pipeline first creates a median, cosmic-ray subtracted flat field image for each mask.  
Wavelength solutions for each slit are fit interactively for the central pixel in
each slit, then propagated outwards to the slit edges to derive a full wavelength grid for each slit. The sky 
background is estimated as a function of wavelength and time using a series of B-splines and subtracted 
from each frame.  The AB frames are differenced, stacked, rectified and output for use along with inverse 
variance-weighted images used for error estimation. Flux calibration was performed using the standard star
on each mask, and agrees with a calibration made using a separate A0V standard star taken during 
twilight to $< 10 \%$.  For our analysis, we assume a 15$\%$ error on the flux calibration in all calculations.

In total we examined 28 targets and secured suitable near-infrared spectra for 20 LBGs in the required redshift range
$3.0<z<3.8$. Of these 13 have pre-existing optical spectra from DEIMOS and 7 represent new MOSFIRE spectroscopic
confirmations determined from our photometric sub-sample (Section 2.3). Table 1 and Fig. \ref{fig:zspec_hist} summarize
the salient properties of the final sample.

\section{Analysis}

The primary goal of this work is to verify or otherwise the conclusions of \citet{Stark2013a} which 
examined the equivalent width distribution for H$\alpha$ emission for LBGs of known spectroscopic
redshift in the range $3.8 < z < 5.0$. A significant conclusion from this study was the remarkably strong emission 
deduced by SED fitting.  Using near-infrared spectroscopy with MOSFIRE we can not only directly measure the
equivalent width distribution of [O III] emission but also test the robustness of the SED-fitting approach by
comparing spectroscopic line fluxes with those inferred from broad-band photometry.

\subsection{Equivalent width distribution}

Even a cursory inspection of our MOSFIRE spectra revealed the presence of many intense line emitters.
Figure 2 shows the 2-D spectra for 4 targets showing strong emission where the continuum remains
undetected. To assemble the equivalent width (EW) distribution, we considered all objects with a spectroscopic detection in
either our DEIMOS or MOSFIRE campaigns (Table 1).  Our sample of 20 objects spans 
$2.97 \leq z \leq 3.77$ with a median of $z = 3.47$. To determine the continuum level necessary to measure
the EW,  we corrected the K-band photometry for the observed fluxes of any emission lines seen in the
MOSFIRE spectra. Where one of the [OIII] doublet lines was partially or fully obscured by a skyline, we assumed
its flux adopting a fixed 5007/4959 line ratio of 3.0. For those spectroscopic targets for which no significant line 
fluxes were detected in our MOSFIRE data, we derived 1-$\sigma$ upper limits for each nebular line EW.  The photometry
for one of our targets, N33\_19374, is likely contaminated by a nearby object.  In this case, we made no attempt
to correct for the flux falling inside our apertures from other objects, but estimate the EW as a lower limit. 

Equivalent widths and errors were measured using a Bayesian Monte Carlo technique.  Because our equivalent
widths depend on the measured line fluxes both directly and indirectly, through the subtraction of the emission
component from the Ks-band photometry, accurate errors are non-trivial and can be asymmetric.  To account
for this, we run a Monte Carlo simulation with N = 10000 trials for each galaxy.  This simulation takes the 
actual measured line fluxes and $K_S$ photometry, perturbs each by the appropriate error, then calculates
the appropriate continuum magnitude and EWs.  At each step, we apply a prior that the fluxes must not be
negative, to ensure a distribution that reflects reality.

Table 2 presents the spectroscopic line measures and the rest-frame EW distribution of [O III] 4959 + 5007 is 
presented in Fig. \ref{fig:ew_dist}.  It is immediately clear that most have very intense emission lines with a median 
[OIII] EW of 280 \AA. There is a significant tail to much higher values;  two galaxies have EWs $>$ 1000 \AA, 
where the $K_S$-band 
photometry is dominated by line emission.  Importantly, we see no significant difference in the distribution
for those galaxies selected on the basis of their DEIMOS spectroscopy and those photometrically
selected entirely for this study.

\subsection{Comparison with Stark et al. 2013}

Although the foregoing suggests that intense line emission sufficient to significantly influence the broad-band photometry 
is quite a common property of $z\simeq$ 3-4 LBGs, we now turn to whether the EW distribution within the 
present sample of 20 LBGs supports the conclusions of \citet{Stark2013a} who derived the EW distribution of $H\alpha$ from a 
larger sample of 45 LBGs with $3.8<z<5.0$ using SED fitting.

Firstly, it is important to determine whether the LBG samples in the two studies are broadly comparable.
In Fig. \ref{fig:stark_comp} we demonstrate that the UV luminosity distribution for the two samples are fairly
similar, with a median absolute magnitude M$_{UV}$ = -21.0 for the \cite{Stark2013a} objects compared
to  M$_{UV}$ = -20.0 for our present sample. Since the UV luminosity correlates closely with the star
formation rate and prominence of Lyman $\alpha$ emission \citep{Stark2010a}, this suggests that
their nebular emission properties should not be too dissimilar.

\cite{Stark2013a} found that the strength of H$\alpha$ emission at $z\simeq$4.5 could be fit well by a 
log normal distribution with log$_{10}$(EW / \AA) = 2.57 and $\sigma = 0.25$.  To facilitate a comparison
with the present [O III] data, we adopt a value of 2.2 for the flux ratio of [OIII] to H$\alpha$, taken from the empirical
compilation of \citep{Anders2003a}.  This ratio is appropriate for a metallicity $Z = 0.2 Z_{\odot}$, consistent with that
inferred from a stack of LBG spectra at $z \sim 4$ \citep{Jones2012a}, as well as measurements of ionized gas in 
LBGs at $z \sim 3.5$ \citep{Maiolino2008a}.

To simulate the expected [O III] EW distribution, we must also account for noise in both the $K_S$-band photometry and the line 
fluxes, which we incorporate using a Monte Carlo distribution using the uncertainties quoted earlier.  
The result is the curve in Fig. \ref{fig:ew_dist} which provides an acceptable fit to the MOSFIRE data and 
a secure confirmation of large nebular line equivalent widths.  We also determine the best fit lognormal
distribution to our measurements of [OIII] EW.  For our total sample of 20 galaxies with either MOSFIRE-confirmed
nebular emission lines, DEIMOS Ly$\alpha$ redshifts, or both, we find best-fit parameters of 
log$_{10}$(EW / \AA) = 2.4 and $\sigma = 0.35$. As this sample has a median redshift $\Delta z \sim 1$
below the \cite{Stark2013a} sample, the slightly weaker than predicted line strengths are not surprising.

Since we also observe the H$\beta$ line in several of our MOSFIRE spectra, this provides an alternative check on
the expected strength H$\alpha$ at these redshifts.  To compute the expected H$\alpha$ flux, 
we adopt Case B recombination and compute the differential reddening determined from 
the best fit SED assuming a Calzetti law
\citep{Calzetti2000a}. The stellar continuum at the location of H$\alpha$ is likewise estimated from the best fit SED.
Taking the average derived H$\alpha$ EW for all 13 DEIMOS galaxies in our sample, we find a value of EW(H$\alpha$)
 = 380 \AA, providing further support for the strong lines derived in \cite{Stark2013a}.

\subsection{Verifying the SED fitting method}

In addition to verifying that our MOSFIRE data on [O III] emission is broadly consistent with the inferences
for H$\alpha$ deduced from SED fitting, we can perform one final check by applying the SED fitting method
used by \citet{Stark2013a} to the present sample and compare the inferred [O III] fluxes with those measured
directly in the near-infrared spectra.

For all objects with either an IRAC 3.6 or 4.5 $\mu$m detection, we predict the [O III] line flux using the 
SED fitting technique adopted by \cite{Stark2013a}.  We fit a grid of Bruzual and Charlot 2003 (BC03) 
stellar continuum models to the observed photometry of each galaxy, excluding the K-band. For the best fitting 
SED, we then compute a synthetic K-band flux, and determine the emission line strength from the residual.
This can only be applied for 8 galaxies from Table 1 for which a detection is available in at least one IRAC filter.

We list the ratios of the SED-predicted flux to that actually observed for the 8 objects in Table 2.  Overall,
the results are in excellent agreement: only one object, N42\_11065 is a catastrophic outlier, with a 
significant [O III] flux implied from the SED method, but with none observed with MOSFIRE..  The remaining 7 
objects all have predicted to observed fluxes within a factor
of 2.5, and 6 are within a factor $\leq 1.6$.  Such an agreement provides a clear validation that the technique
we used in \cite{Stark2013a} can provide line strength measurements suitable for statistical purposes.

\subsection{Ly$\alpha$ velocity offsets}

Since we now possess sample of $z\simeq$ 3.5 galaxies with both optical and near-infrared spectra,
we can comment briefly on the prevalence of outflows. As is well known, a velocity offset is often observed between 
Ly$\alpha$, which is easily resonantly scattered by hydrogen on its way out of a galaxy, and other nebular lines which
trace directly the sites of star formation and provide a systemic redshift, \citep[e.g.,][]{Shapley2003a}.  
In particular, Ly$\alpha$ is often observed with a positive
velocity offset, suggesting those photons are only able to escape after being scattered by an 
outflowing HI wind on the far side of the galaxy and shifting out of resonance with any
HI on the near side \citep{Steidel2010a}.

Measures of this offset velocity at high-redshift can shed light on some outstanding issues related to cosmic reionization.  
Firstly, one of the major currently unknown variables that enters into reionization calculations remains \fesc, the escape 
fraction of ionizing photons from galaxies.  Although recent measurements have found \fesc $\simeq 10 \%$ at 
$z \sim 3$ \citep{Nestor2013a}, direct measurements are impossible at higher redshifts, owing to the increased opacity
of the intergalactic medium (IGM).  A higher value ($\simeq 20\%$) is required to reproduce measurements of the IGM 
neutral fraction at high redshifts \citep[e.g.,][]{Robertson2013a,Kuhlen2012a}. As any outflowing neutral gas will serve to 
extinguish ionizing radiation, an observed decrease in the velocity offset of Ly$\alpha$, potentially indicating a lower 
covering fraction of neutral gas, would provide further support for an increased \fesc. The velocity offset is also of 
direct interest, as numerous experiments seeking to directly probe the ionization state of the IGM at $z > 6$ utilize the 
visibility of Ly$\alpha$ emission \citep[e.g.,][]{Pentericci2011a,Schenker2012a,Ono2012a}.  If Ly$\alpha$ escapes galaxies
with a smaller velocity offset than previously believed, it is closer to resonance and more easily quenched by a neutral IGM.

To this end, we present the difference in the observed velocities of Ly$\alpha$ and the H$\beta$+[OIII] for 
all objects with at least a 3$\sigma$ line detection in Table 2 and Fig. \ref{fig:velocity}.  The left panel
displays both stacked H$\beta$+[OIII] (black) and Ly$\alpha$ (red) profiles, demonstrating the high-fidelity velocity measurements we are able to make with MOSFIRE.  In the right panel, we present a compilation of Ly$\alpha$
velocity offset measurements, plotted as a function of redshift and Ly$\alpha$ EW.  

As all of the 9 targets in our
MOSFIRE sample show significant Ly$\alpha$ in emission (EW $> 20 $\AA), we must be careful not to draw
conclusions by blindly comparing this to the \cite{Steidel2010a} sample at lower redshift, for which all objects
have only modest equivalent widths.  A more illuminating conclusion can perhaps be drawn by compiling
the velocity offsets for other galaxies with strong Ly$\alpha$ emission, drawn here from 
\cite{McLinden2011a,Finkelstein2011a,Hashimoto2013a}.  All have quite low offsets, with 
$<v_{Ly\alpha}> = +149$ km s$^{-1}$ for the entire sample of 17 objects, and $<v_{Ly\alpha}> =+157$ km s$^{-1}$
for our own 9.  As the fraction of starforming galaxies displaying strong Ly$\alpha$ emission increases with redshift
out to at least $z \sim 6$ \citep{Stark2010a}, this data implies it may become easier for ionizing photons
to escape if this correlation of large EW$_{Ly\alpha}$ with $v_{Ly\alpha}$ trend is in fact caused by lower HI covering fractions. 

\section{Discussion}

We have shown, through near-infrared spectroscopy with MOSFIRE, that the main conclusions of our
earlier work \citep{Stark2013a} are confirmed. A significant fraction of our 20 LBG targets show
intense [O III] line emission and the EW distribution is broadly comparable with that inferred for H$\alpha$
from SED fitting for a larger sample. Moreover, where we can make a direct comparison within our
own sample, the SED fitting method predicts [O III] line fluxes that are in reasonable agreement,
given the uncertainties, with those measured directly with MOSFIRE.

One might worry that because our DEIMOS-confirmed objects mostly display large Ly$\alpha$ equivalent widths 
(a result of following up only secure confirmations with MOSFIRE) that we are biasing our sample toward 
especially strong emitters.  However, our additional photometric sample of 7 objects dissuades this notion.
Although the samples remain modest in size, not only does the EW distribution appear similar (Fig. \ref{fig:ew_dist}), but 
our largest equivalent width object, N33\_18453, is also part of this photometric sample.

We can also examine, for our present sample, how our measured line emission affects the derived physical properties. 
Fig. \ref{fig:seds} presents SEDs for a selection of our sample where the excess flux in the $K_S$ band
is clearly visible. We can fit the SEDs using both the entire photometric dataset, including the line-contaminated Ks filter, 
as well as that excluding the contaminated band.  In both cases, we assume a stellar continuum only. 
In this trial, the median properties of the sample hardly change; there is no significant change in age,
and the stellar mass is reduced by only 3$\%$ when correcting for line emission. This is because the IRAC 3.6 and 
4.5 $\mu$m photometry provides a crucial measurement free from line contamination longward of the Balmer break.  
Thus the majority of existing measurements of SED-derived properties of LBGs at $z \sim 3$ which incorporate
IRAC data, should not be significantly affected by [O III] emission, even though it is particularly intense.

A more illuminating test applies when the IRAC photometry is ignored, in which case the $K_S$ filter becomes
the only photometric measurement beyond the Balmer break.  This is a more appropriate test of how SEDs are 
fit at z $\sim$ 6-7, where both the 3.6 and 4.5 $\mu$m IRAC filters are contaminated by [OIII] and H$\alpha$, respectively.  
In this comparison, the implications of line emission are much more striking.  The median mass for the line corrected SEDs 
is only 64\% that of the mass determined using the contaminated photometry, and the median age is lowered by 30$\%$.  
For the most intense emitters, N33\_24311 and N33\_18453, the masses can be reduced by factors of $\simeq 20$.

We have demonstrated here, for the first time with both robust spectroscopy and significant sample sizes, 
the strength of nebular emission in $z \geq 3$ LBGs, as well as significantly increased the number of the
same galaxies with measurements of Ly$\alpha$ velocity offsets.  The implications of such observations are 
extremely important for the high redshift universe.  SED fitting at high redshifts must account for 
contamination of broadband filters by these strong lines to determine accurate stellar masses.  These lines
provide mounting evidence for a continued increase of the sSFR beyond $z = 2$, which has only 
recently been suggested.  The measurements of precise offsets also provide valuable input to models
which seek to map out the universal neutral fraction through Ly$\alpha$ radiative transfer, and also 
bolster arguments for an increasing escape fraction with redshift.  With the era of multi-object, near-infrared 
spectrographs just now beginning, prospects for further solidification of these trends will be strong.

\acknowledgements

We thank Chuck Steidel and Ian McLain for their hard work in developing the MOSFIRE 
instrument.  The Keck observatory staff proved invaluable, and we thank them for their dedication to 
maintaining a world-class observatory.  We also wish
to recognize and acknowledge the very signiﬁcant cultural role and reverence that the summit of Mauna Kea
has always had within the indigenous Hawaiian community. We are most fortunate to have the opportunity to
conduct observations from this mountain.

\begin{deluxetable*}{lcccccccccc}
\tablecolumns{10}
\tablewidth{0pt}
\tablecaption{\bf Properties of observed targets}
\tablehead{\colhead{ID} & \colhead{RA} & \colhead{Dec} & \colhead{$m_{z850}$} & \colhead{$z_{spec}$} 
& \colhead{Mass / [$10^9 M_{\odot}$]} & \colhead{SFR [$M_{\odot}$ yr$^{-1}$] }
& \colhead{E(B-V)} & \colhead{log$_{10}$(age / yr)} & \colhead{m$_{Ks}$\tablenotemark{b}}}
\medskip
\startdata

N33\_24311 & 12:37:06.55 & 62:15:35.4 & 25.3 & 3.474 & 0.8 & 29.7 & 0.10 & 7.48 & $24.1 \pm 0.2$  \\
N33\_19880 & 12:36:55.14 & 62:15:29.2 & 26.7 & 3.405 & 2.3 & 2.1 & 0.00 & 9.16 & $25.1 \pm 0.4$  \\
N33\_25713 & 12:37:10.62 & 62:14:52.6 & 25.6 & 3.615 & 5.7 & 37.1 & 0.20 & 8.26 & $24.5 \pm 0.2$  \\
N33\_18549 & 12:36:51.89 & 62:15:14.5 & 26.0 & 3.333 & 7.7 & 13.5 & 0.15 & 8.86 & $24.3 \pm 0.2$  \\
N33\_19374 & 12:36:53.88 & 62:14:18.0 & 26.3 & 3.653 & 75.7 & 68.2 & 0.35 & 9.16 & $23.9 \pm 0.1$  \\
N33\_24278 & 12:37:06.46 & 62:13:20.5 & 25.8 & 3.671 & 3.7 & 12.3 & 0.10 & 8.56 & $25.2 \pm 0.5$  \\
N33\_25726 & 12:37:10.66 & 62:12:39.0 & 25.2 & 3.733 & 11.5 & 38.6 & 0.15 & 8.56 & $24.0 \pm 0.1$  \\
N32\_20647 & 12:36:57.00 & 62:11:51.0 & 26.3 & 3.740 & 1.1 & 9.2 & 0.10 & 8.16 & $25.6 \pm 0.9$  \\
N32\_23933 & 12:37:05.52 & 62:11:27.2 & 25.3 & 3.469 & 6.8 & 696.1\tablenotemark{a} & 0.45 & 7.00 & $23.4 \pm 0.1$  \\
N32\_15359 & 12:36:43.54 & 62:11:21.4 & 24.4 & 3.488 & 11.1 & 1139.7\tablenotemark{a} & 0.40 & 7.00 & $23.0 \pm 0.1$  \\
N42\_7697 & 12:36:22.17 & 62:09:42.3 & 25.8 & 3.771 & 1.2 & 18.3 & 0.10 & 7.86 & $25.5 \pm 0.7$  \\
N42\_12130 & 12:36:35.15 & 62:08:50.8 & 26.0 & 3.474 & 1.7 & 79.6 & 0.30 & 7.36 & $25.5 \pm 0.7$  \\
N42\_11065 & 12:36:32.24 & 62:09:46.5 & 25.4 & 3.658 & 1.4 & 39.5 & 0.15 & 7.59 & $25.1 \pm 0.4$  \\
N32\_14225 & 12:36:40.58 & 62:10:41.1 & 26.0 & 3.245 & 5.8 & 89.9 & 0.35 & 7.86 & $24.1 \pm 0.2$  \\
N32\_16805 & 12:36:47.57 & 62:10:23.7 & 25.9 & 3.235 & 7.2 & 65.1 & 0.30 & 8.11 & $23.9 \pm 0.1$  \\
N32\_15430 & 12:36:43.79 & 62:11:20.0 & 25.8 & 3.237 & 5.1 & 124.8 & 0.35 & 7.65 & $23.9 \pm 0.1$  \\
N32\_19795 & 12:36:54.94 & 62:11:43.8 & 25.1 & 2.976 & 16.1 & 162.0 & 0.35 & 8.06 & $23.2 \pm 0.1$  \\
N33\_23907 & 12:37:05.48 & 62:12:37.1 & 25.8 & 3.678 & 2.0 & 11.8 & 0.10 & 8.31 & $25.1 \pm 0.4$  \\
N33\_20428 & 12:36:56.48 & 62:13:39.9 & 26.8 & 3.436 & 3.4 & 7.5 & 0.15 & 8.76 & $24.6 \pm 0.3$  \\
N33\_18453 & 12:36:51.69 & 62:15:10.1 & 25.8 & 3.364 & 0.2 & 25.5 & 0.10 & 7.00 & $24.4 \pm 0.2$  \\

 \smallskip
\enddata
\tablecomments{Sample IDs, coordinates, redshifts, and properties derived from SED fitting.  Where
appropriate, emission line flux has been subtracted from the $K_S$-band photometry before
fitting to the photometry, using only the stellar continuum to determine SED properties.}
\tablenotetext{a}{When SFRs for these objects are estimated using the dust correction from
\cite{Meurer1999a} and the UV continuum to SFR conversion of \cite{Kennicutt1998a}, we find
much more modest rates of 30 and 150 $M_{\odot}$ yr$^{-1}$, respectively.}
\tablenotetext{b}{Total measured $K_S$ magnitude, including any line emission.}

\end{deluxetable*}

\begin{deluxetable*}{lcccccccccc}
\tablecolumns{10}
\tablewidth{0pt}
\tablecaption{\bf MOSFIRE spectroscopic measurements}
\tablehead{\colhead{ID} & \colhead{flux$_{H\beta}$ [$10^{-18}$ erg/cm/s]} &
\colhead{flux$_{[OIII]}$ [$10^{-18}$ erg/cm/s]}  & \colhead{EW$_{H\beta}$ [\AA, rest]} &
\colhead{EW$_{[OIII]}$ [\AA, rest]} & \colhead{$v_{neb} - v_{Ly\alpha}$ [km/s]\tablenotemark{a}} &
\colhead{$\frac{flux_{pred,SED}}{flux_{obs}}$} }
\medskip
\startdata

N33\_24311 & 13.5 $\pm$ 2.0 & 99.5 $\pm$ 3.2 & $140^{+240}_{-100}$  & $1100^{+600}_{-200}$  & 152 & 0.91 \\
N33\_19880 & 5.9 $\pm$ 1.7 & 35.9 $\pm$ 3.9 & $89^{+196}_{-35}$  & $600^{+550}_{-200}$  & 4 &  ...  \\
N33\_25713 & 7.2 $\pm$ 2.3 & 21.7 $\pm$ 4.2 & $47^{+78}_{-29}$  & $150^{+70}_{-30}$  & 106 &  ...  \\
N33\_18549 & 22.5 $\pm$ 2.9 & 38.9 $\pm$ 2.5 & $170^{+240}_{-120}$  & $300^{+100}_{-50}$  & 319 & 0.74 \\
N33\_19374 & -3.0 $\pm$ 2.7 & 1.1 $\pm$ 4.7 & $0^{+7}_{-0}$  & $0^{+21}_{-0}$  & xxx &  ...  \\
N33\_24278 & -8.6 $\pm$ 3.2 & 32.8 $\pm$ 6.4 & $0^{+20}_{-0}$  & $370^{+190}_{-120}$  & 207 &  ...  \\
N33\_25726 & 9.5 $\pm$ 3.8 & 79.9 $\pm$ 8.0 & $40^{+61}_{-21}$  & $360^{+60}_{-50}$  & 86 &  ...  \\
N32\_20647 & 9.8 $\pm$ 3.6 & 56.2 $\pm$ 7.3 & $130^{+310}_{-40}$  & $890^{+690}_{-320}$  & 113 &  ...  \\
N32\_23933 & 14.0 $\pm$ 2.5 & 49.1 $\pm$ 3.8 & $40^{+49}_{-30}$  & $150^{+20}_{-20}$  & 195 & 2.31 \\
N32\_15359 & 11.9 $\pm$ 2.1 & 54.8 $\pm$ 5.0 & $21^{+26}_{-17}$  & $100^{+10}_{-10}$  & ... & 1.39 \\
N42\_7697 & 7.9 $\pm$ 6.2 & 18.6 $\pm$ 9.2 & $91^{+226}_{-5}$  & $260^{+220}_{-160}$  & 227 &  ...  \\
N42\_12130 & 5.5 $\pm$ 3.3 & 6.0 $\pm$ 4.4 & $65^{+196}_{-7}$  & $150^{+210}_{-90}$  & xxx &  ...  \\
N42\_11065 & 2.0 $\pm$ 4.3 & 1.2 $\pm$ 7.3 & $4^{+80}_{-1}$  & $6^{+113}_{-3}$  & xxx & 18.74 \\
N32\_14225 & 7.2 $\pm$ 3.5 & 47.6 $\pm$ 9.0 & $32^{+65}_{-18}$  & $280^{+100}_{-80}$  & ... &  ...  \\
N32\_16805 & 20.7 $\pm$ 12.0 & 39.6 $\pm$ 5.8 & $79^{+162}_{-29}$  & $190^{+50}_{-40}$  & ... & 1.42 \\
N32\_15430 & 11.3 $\pm$ 3.7 & 68.6 $\pm$ 5.2 & $50^{+82}_{-31}$  & $340^{+80}_{-50}$  & ... & 1.01 \\
N32\_19795 & 8.4 $\pm$ 6.9 & 94.2 $\pm$ 7.5 & $17^{+29}_{-4}$  & $190^{+20}_{-20}$  & ... & 1.60 \\
N33\_23907 & 6.0 $\pm$ 4.9 & 18.3 $\pm$ 6.3 & $54^{+139}_{-3}$  & $210^{+200}_{-120}$  & xxx &  ...  \\
N33\_20428 & 8.1 $\pm$ 3.2 & 45.6 $\pm$ 4.9 & $84^{+164}_{-32}$  & $500^{+280}_{-130}$  & ... &  ...  \\
N33\_18453 & 9.6 $\pm$ 1.7 & 126.5 $\pm$ 3.9 & $320^{+860}_{-150}$  & $4400^{+6200}_{-2000}$  & ... & 1.63 \\

\smallskip
\enddata

\tablecomments{MOSFIRE emission line measurements for our sample.}
\tablenotetext{a} {'xxx' denotes DEIMOS objects for which the MOSFIRE emission lines were not robust enough
to permit a velocity offset measurement. '...' denotes objects from our photometric sample.}

\end{deluxetable*}

\begin{figure*}
\begin{center}

\includegraphics[width=0.6\textwidth]{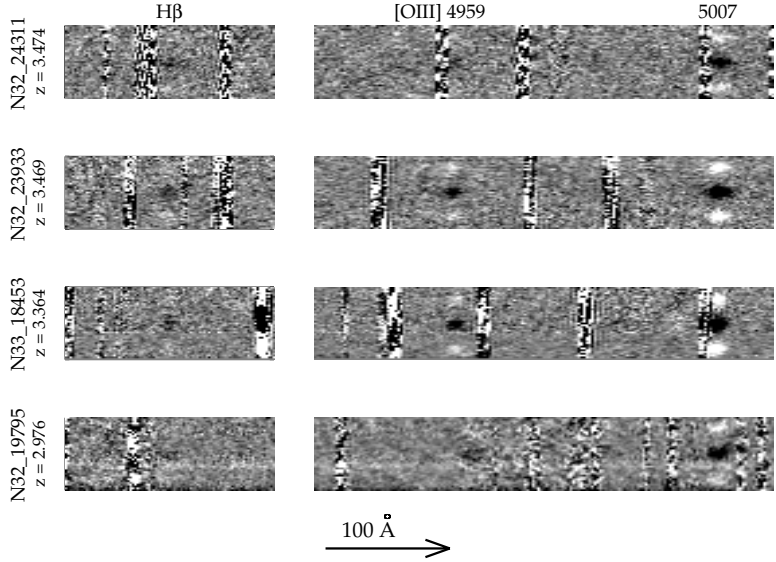}

\caption{\label{fig:spectra} Example 2-D MOSFIRE spectra in A-B-A format for 4 targets
showing prominent nebular emission. The left panels focus on the region
containing H$\beta$ and the right panels the [O III] doublet (marked).}
\end{center}
\end{figure*}

\begin{figure*}
\begin{center}

\includegraphics[width=0.95\textwidth]{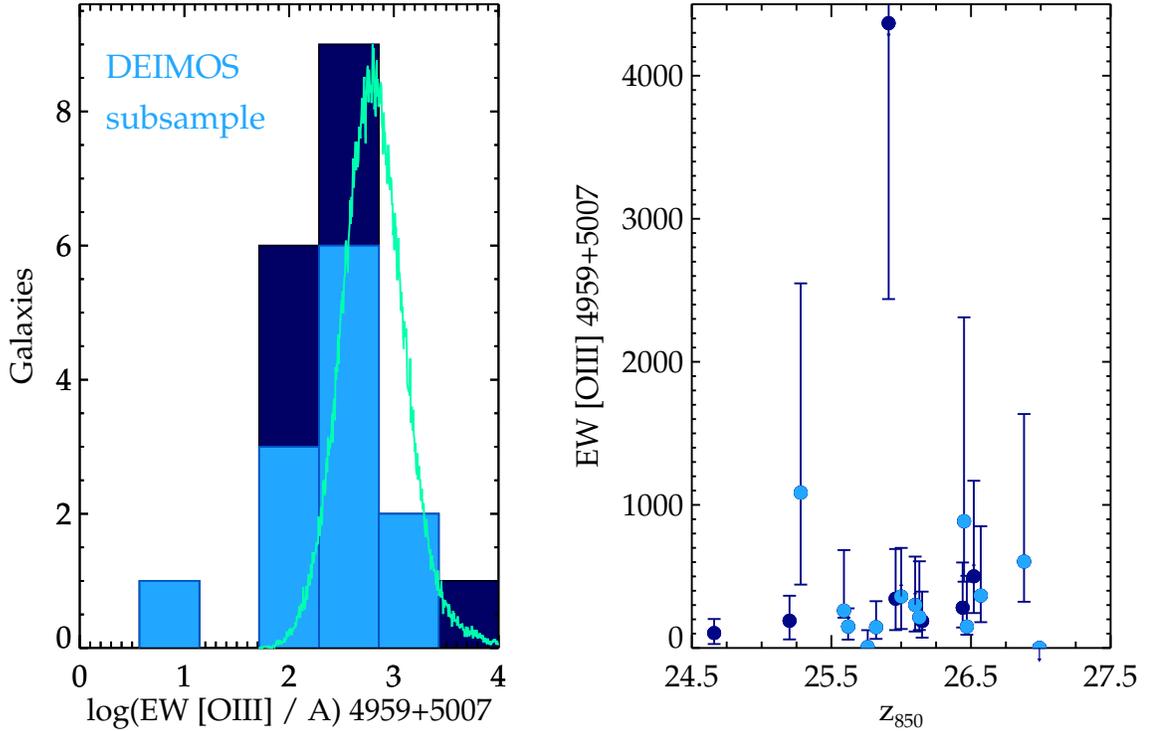}

\caption{\label{fig:ew_dist} Left: The rest-frame equivalent width distribution of [OIII] 4959 + 5007 \AA\ 
derived from our MOSFIRE spectroscopic data. The green curve estimates the distribution expected from
the distribution for H$\alpha$ derived from SED fitting method by \citet{Stark2013a}. Right: Individual equivalent
widths of [O III] versus $z_{850}$ magnitude. In both panels, light blue denotes the spectroscopic sample
and dark blue the photometric sample (see Section 2).}
\end{center}
\end{figure*}

\begin{figure*}
\begin{center}

\includegraphics[width=0.6\textwidth]{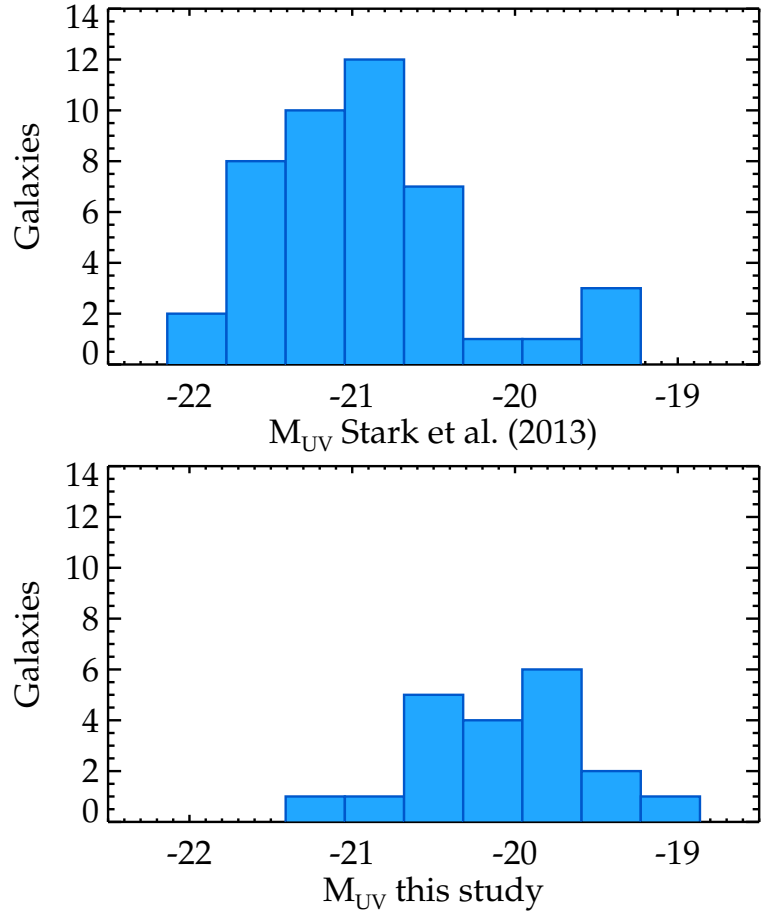}

\caption{\label{fig:stark_comp} Absolute UV magnitude distribution for the $3.8<z<5.0$ sample discussed by \cite{Stark2013a}
(top panel) compared to that at $3.0<z<$3.8 studied in this paper.}
\end{center}
\end{figure*}

\begin{figure*}
\begin{center}

\includegraphics[width=0.45\textwidth]{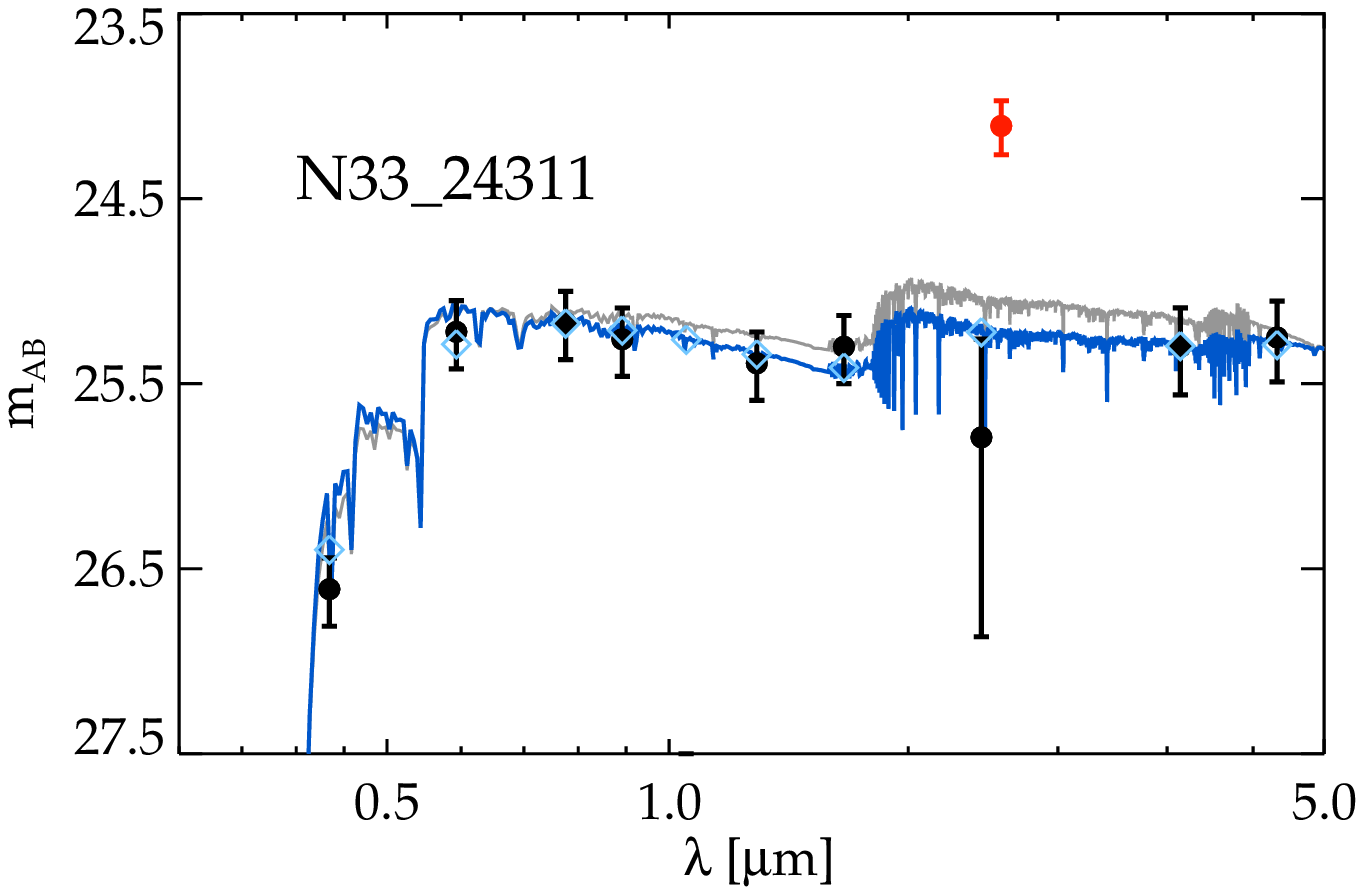}
\includegraphics[width=0.45\textwidth]{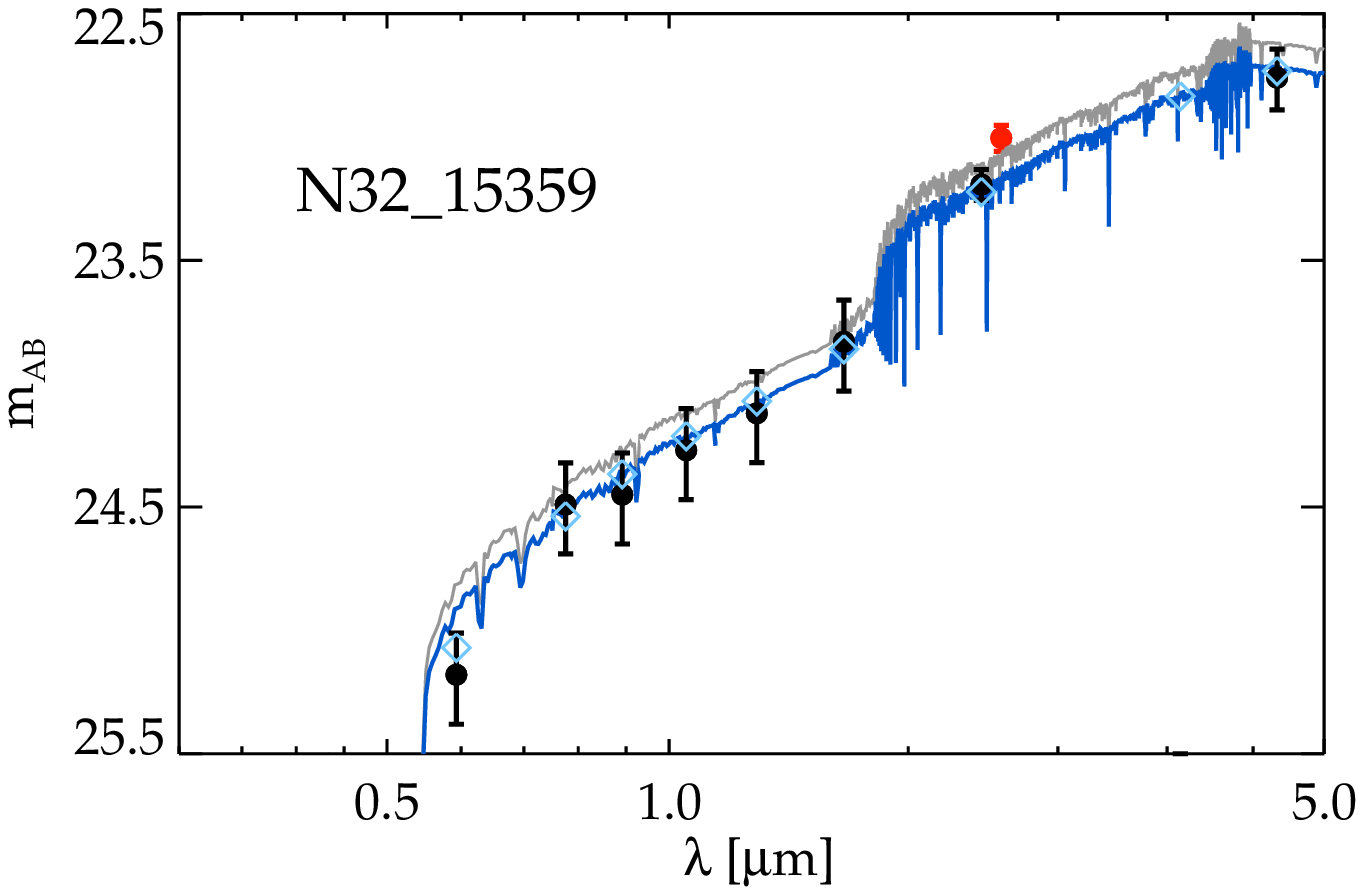}
\includegraphics[width=0.45\textwidth]{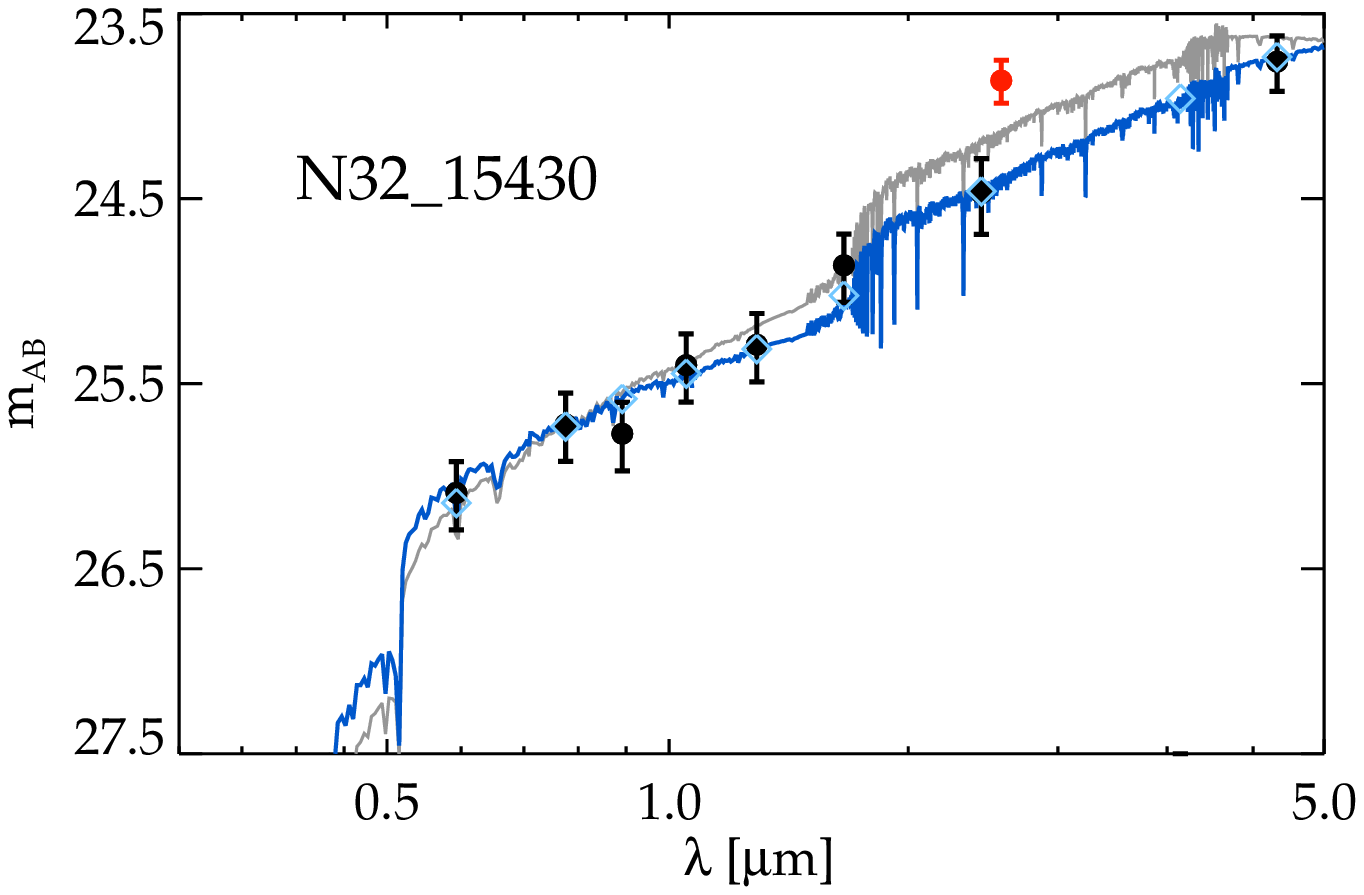}

\caption{\label{fig:seds} Spectral energy distributions for three of our targets.  The red data point represents the observed K-band
photometry without correction for [O III] contamination, and the grey spectrum shows the best fit SED to this data. The black K-band
data point shows stellar continuum flux after correction for the MOSFIRE-determined [O III] line flux, and the blue spectrum
is the associated best fit SED.}
\end{center}
\end{figure*}

\begin{figure*}
\begin{center}

\includegraphics[width=0.35\textwidth]{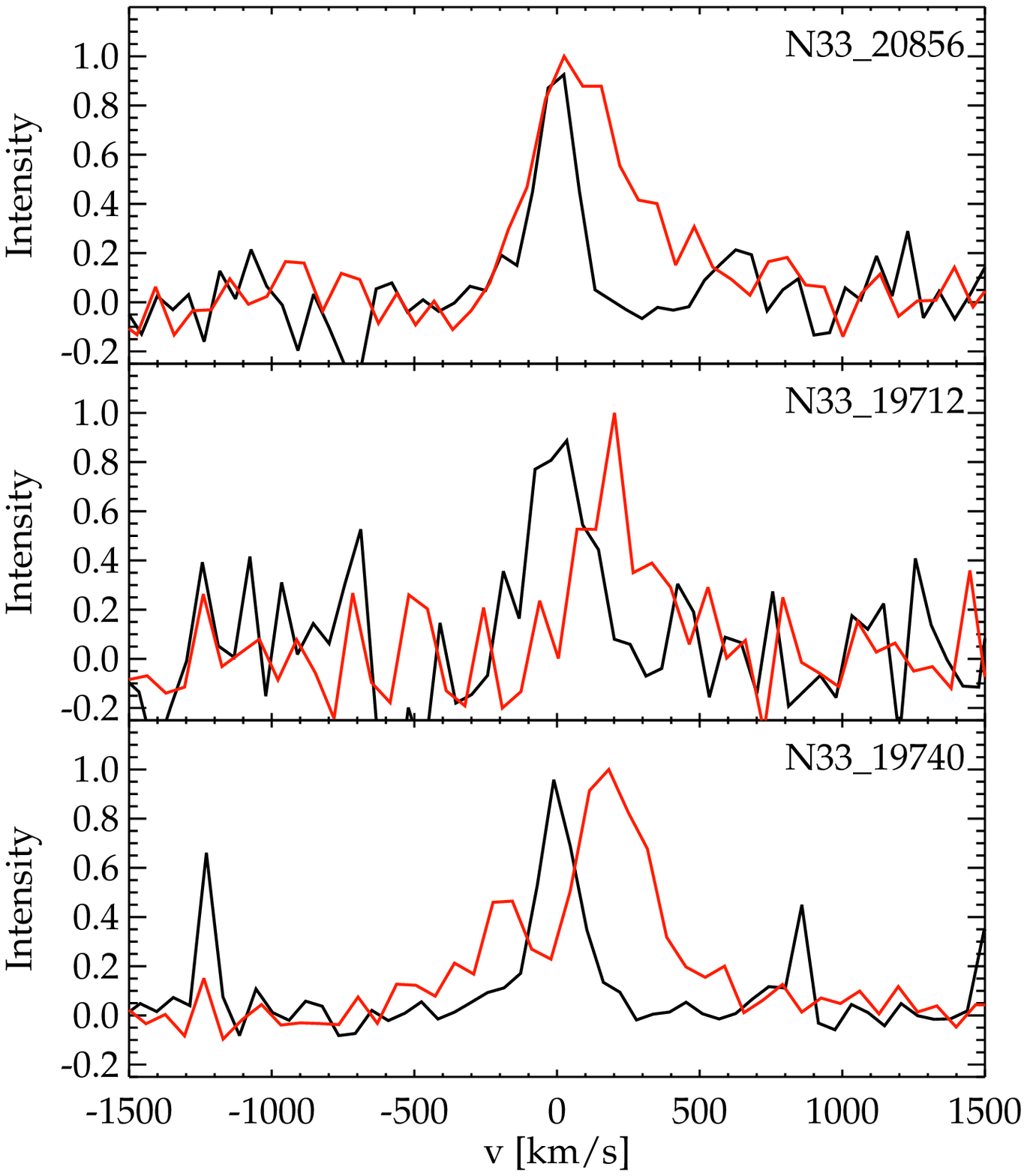}
\includegraphics[width=0.55\textwidth]{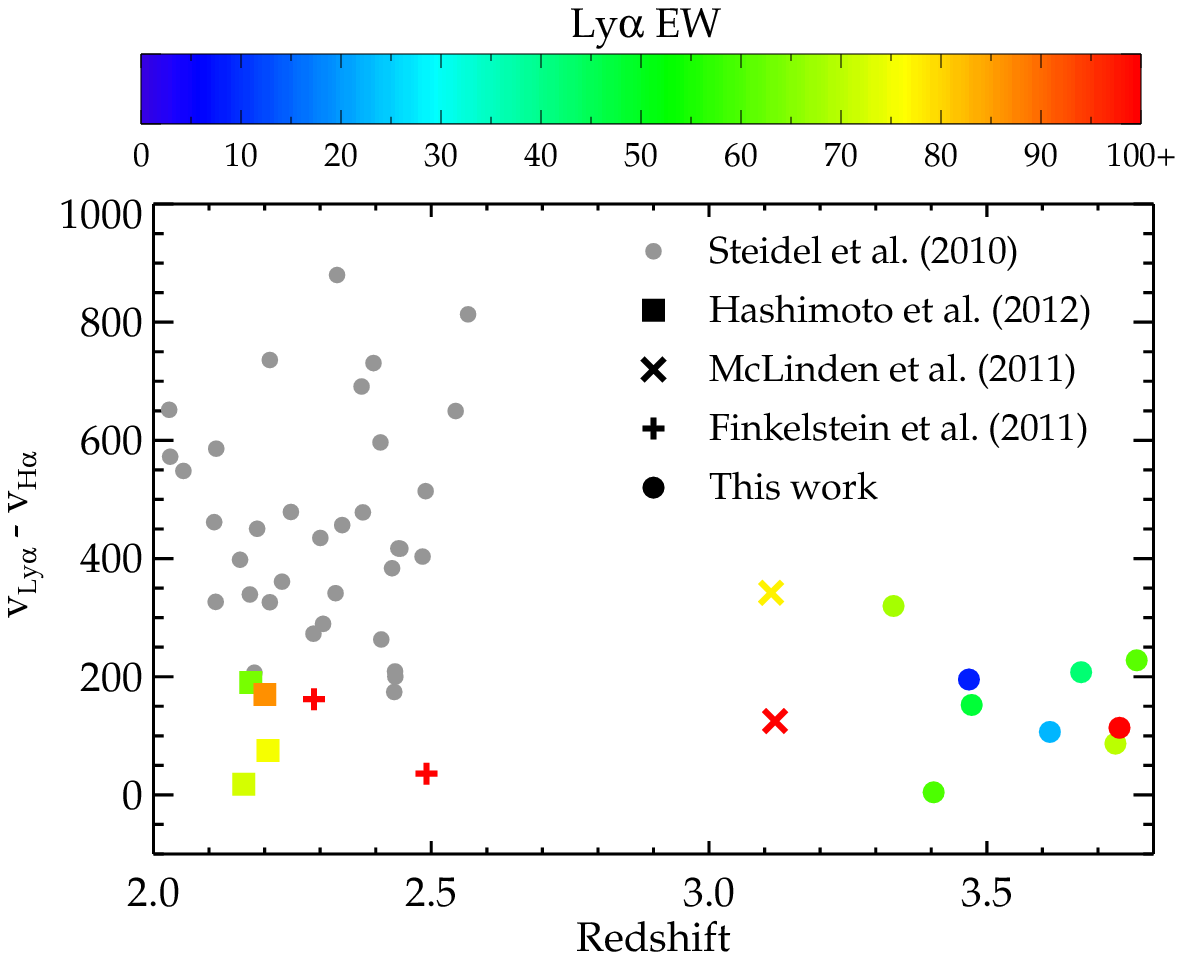}

\caption{\label{fig:velocity} Left: The velocity structure observed by comparing MOSFIRE and DEIMOS spectra for three objects
from our spectroscopic sample. Zero velocity is defined using a stack of the nebular [OIII] and H$_{\beta}$ lines, shown in 
black.  Ly$\alpha$ is overlaid in red, arbitrarily scaled in the y-axis.  Right: Compilation of Ly$\alpha$ velocity
offset measurements from various sources in the literature. Our objects show much more modest offsets than
the sample of \cite{Steidel2010a}, which all display only low level Ly$\alpha$ emission, but are more consistent
with numerous LAEs from the literature, also plotted.}
\end{center}
\end{figure*}


\medskip

\begin{thebibliography}{38}
\expandafter\ifx\csname natexlab\endcsname\relax\def\natexlab#1{#1}\fi


\bibitem[Anders 
\& Fritze-v.~Alvensleben(2003)]{Anders2003a} Anders, P., \& Fritze-v.~Alvensleben, U.\ 2003, \aap, 401, 1063 


\bibitem[Bertin et al.(2002)]{Bertin2002a} Bertin, E., Mellier, Y., 
Radovich, M., Missonnier, G., Didelon, P., 
\& Morin, B.\ 2002, Astronomical Data Analysis Software and Systems XI, 281, 228 


\bibitem[Calzetti et al.(2000)]{Calzetti2000a} Calzetti, D., Armus, L., 
Bohlin, R.~C., Kinney, A.~L., Koornneef, J., 
\& Storchi-Bergmann, T.\ 2000, \apj, 533, 682

\bibitem[Dav{\'e}, Oppenheimer, 
\& Finlator(2011)]{Dave2011a} Dav{\'e}, R., Oppenheimer, B.~D., \& Finlator, K.\ 2011, \mnras, 415, 11 

\bibitem[Dav{\'e}, Finlator \& Oppenheimer (2012a)]{Dave2012a}
Dav{\'e}, R.,  Finlator, K.  \& Oppenheimer, B.~D. \ 2012, \mnras, 421, 98 


\bibitem[de Barros, Schaerer, 
\& Stark(2012)]{deBarros2012a} de Barros, S., Schaerer, D., \& Stark, D.~P.\ 2012, arXiv:1207.3663 


\bibitem[Faber et al.(2003)]{Faber2003a} Faber, S.~M., et al.\ 2003, 
\procspie, 4841, 1657 

\bibitem[Finkelstein et al.(2011)]{Finkelstein2011a} Finkelstein, S.~L., et 
al.\ 2011, \apj, 729, 140 


\bibitem[Giavalisco et al.(2004)]{Giavalisco2004a} Giavalisco, M., et al.\ 
2004, \apjl, 600, L93 

\bibitem[Gonz{\'a}lez et al.(2010)]{Gonzalez2010a} Gonz{\'a}lez, V., 
Labb{\'e}, I., Bouwens, R.~J., Illingworth, G., Franx, M., Kriek, M., 
\& Brammer, G.~B.\ 2010, \apj, 713, 115 

\bibitem[Gonz{\'a}lez et al.(2011)]{Gonzalez2011a} Gonz{\'a}lez, V., 
Labb{\'e}, I., Bouwens, R.~J., Illingworth, G., Franx, M., 
\& Kriek, M.\ 2011, \apjl, 735, L34 

\bibitem[Gonz{\'a}lez et al.(2012)]{Gonzalez2012a} Gonz{\'a}lez, V., Bouwens, R., 
llingworth, G., Labbe, I., Oesch, P., Franx, M., 
\& Magee, D.\ 2012, arXiv:1208.4362 

\bibitem[Grogin et al.(2011)]{Grogin2011a} Grogin, N.~A., et al.\ 2011, 
\apjs, 197, 35 

\bibitem[Hashimoto et al.(2013)]{Hashimoto2013a} Hashimoto, T., Ouchi, M., 
Shimasaku, K., Ono, Y., Nakajima, K., Rauch, M., Lee, J., 
\& Okamura, S.\ 2013, \apj, 765, 70 


\bibitem[Jones, Stark, 
\& Ellis(2012)]{Jones2012a} Jones, T., Stark, D.~P., \& Ellis, R.~S.\ 2012, \apj, 751, 51 


\bibitem[Kelson(2003)]{Kelson2003a} Kelson, D.~D.\ 2003, \pasp, 115, 688 

\bibitem[Kennicutt(1998)]{Kennicutt1998a} Kennicutt, R.~C., Jr.\ 1998, 
\araa, 36, 189 

\bibitem[Kuhlen 
\& Faucher-Gigu{\`e}re(2012)]{Kuhlen2012a} Kuhlen, M., \& Faucher-Gigu{\`e}re, C.-A.\ 2012, \mnras, 423, 862

\bibitem[Koekemoer et al.(2011)]{Koekemoer2011a} Koekemoer, A.~M., et al.\ 
2011, \apjs, 197, 36 

\bibitem[Labb{\'e} et al.(2010)]{Labbe2010a} Labb{\'e}, I., et al.\ 2010, 
\apjl, 716, L103 

\bibitem[Maiolino et al.(2008)]{Maiolino2008a} Maiolino, R., et al.\ 2008, 
\aap, 488, 463

\bibitem[McLean et al.(2012)]{McLean2012a} McLean, I.~S., et al.\ 2012, 
\procspie, 8446

\bibitem[McLinden et al.(2011)]{McLinden2011a} McLinden, E.~M., et al.\ 
2011, \apj, 730, 136 

\bibitem[Meurer, Heckman, 
\& Calzetti(1999)]{Meurer1999a} Meurer, G.~R., Heckman, T.~M., \& Calzetti, D.\ 1999, \apj, 521, 64 

\bibitem[Nestor et al.(2013)]{Nestor2013a} Nestor, D.~B., Shapley, A.~E., 
Kornei, K.~A., Steidel, C.~C., \& Siana, B.\ 2013, \apj, 765, 47 


\bibitem[Oke 
\& Gunn(1983)]{Oke1983a} Oke, J.~B., \& Gunn, J.~E.\ 1983, \apj, 266, 713 


\bibitem[Ono et al.(2010)]{Ono2010a} Ono, Y., Ouchi, M., Shimasaku, K., 
Dunlop, J., Farrah, D., McLure, R., \& Okamura, S.\ 2010, \apj, 724, 1524 

\bibitem[Ono et al.(2012)]{Ono2012a} Ono, Y., et al.\ 2012, \apj, 744, 83 

\bibitem[Pentericci et al.(2011)]{Pentericci2011a} Pentericci, L., et al.\ 
2011, \apj, 743, 132 

\bibitem[Robertson et al.(2013)]{Robertson2013a} Robertson, B.~E., et al.\ 
2013, \apj, 768, 71 


\bibitem[Shapley et al.(2003)]{Shapley2003a} Shapley, A.~E., Steidel, C.~C., Pettini, M., \& Adelberger, K.~L.\ 2003, \apj, 588, 65 


\bibitem[Schaerer 
\& de Barros(2010)]{Schaerer2010a} Schaerer, D., \& de Barros, S.\ 2010, \aap, 515, A73 


\bibitem[Schaerer 
\& de Barros(2009)]{Schaerer2009a} Schaerer, D., \& de Barros, S.\ 2009, \aap, 502, 423 


\bibitem[Schenker et al.(2012)]{Schenker2012a} Schenker, M.~A., Stark, 
D.~P., Ellis, R.~S., Robertson, B.~E., Dunlop, J.~S., McLure, R.~J., Kneib, 
J.-P., \& Richard, J.\ 2012, \apj, 744, 179 

\bibitem[Shim et al.(2011)]{Shim2011a} Shim, H., Chary, R.-R., Dickinson, 
M., Lin, L., Spinrad, H., Stern, D., \& Yan, C.-H.\ 2011, \apj, 738, 69 


\bibitem[Stark et al.(2009)]{Stark2009a} Stark, D.~P., Ellis, R.~S., 
Bunker, A., Bundy, K., Targett, T., Benson, A., 
\& Lacy, M.\ 2009, \apj, 697, 1493 


\bibitem[Stark et al.(2010)]{Stark2010a} Stark, D.~P., Ellis, R.~S., Chiu, 
K., Ouchi, M., \& Bunker, A.\ 2010, \mnras, 408, 1628 


\bibitem[Stark, Ellis, 
\& Ouchi(2011)]{Stark2011a} Stark, D.~P., Ellis, R.~S., \& Ouchi, M.\ 2011, \apjl, 728, L2 

\bibitem[Stark et al.(2013)]{Stark2013a} Stark, D.~P., Schenker, M.~A., 
Ellis, R., Robertson, B., McLure, R., \& Dunlop, J.\ 2013, \apj, 763, 129 

\bibitem[Steidel et al.(2010)]{Steidel2010a} Steidel, C.~C., Erb, D.~K., 
Shapley, A.~E., Pettini, M., Reddy, N., Bogosavljevi{\'c}, M., Rudie, 
G.~C., \& Rakic, O.\ 2010, \apj, 717, 289 


\bibitem[Wang et al.(2010)]{Wang2010a} Wang, W.-H., Cowie, L.~L., Barger, 
A.~J., Keenan, R.~C., \& Ting, H.-C.\ 2010, \apjs, 187, 251 


\end{thebibliography}
\end{document}